# Self-consistent simulation of plasma scenarios for ITER using a combination of 1.5D transport codes and free boundary equilibrium codes

V Parail 1), R Albanese 2), R Ambrosino 2), J-F Artaud 3), K Besseghir 4), M Cavinato 5), G Corrigan 1), J Garcia 3), L Garzotti 1), Y Gribov 6), F Imbeaux 3), F Koechl 7), CV Labate 2), J Lister 4), X Litaudon 3), A Loarte 6), P Maget 3), M Mattei 2), D McDonald 1,8), E Nardon 3), G Saibene 5), R Sartori 5) and J. Urban 3,9)
1) Euratom/CCFE Fusion Association, Culham Science Centre, Abingdon, OX14 3DB, UK
2) Association EURATOM-ENEA-CREATE, Italy
3) CEA, IRFM, F-13108 St Paul-Lez-Durance, France
4) Association EURATOM-Confederation Suisse, CRPP, EFPL, Lausanne, Switzerland
5) Fusion for Energy, Barcelona, Spain
6) ITER Organization, Route de Vinon sur Verdon, 13115 St Paul Lez Durance, France
7) Association EURATOM-ÖAW/ATI, Atominstitut, TU Wien, 1020 Vienna, Austria
8) EFDA CSU Garching, Boltzmannstr. 2, 85748 Garching bei München, Germany
9) Institute of Plasma Physics AS CR, v.v.i., Association EURATOM/IPP.CR, Prague, Czech Republic
e-mail address: vassili.parail@ccfe.ac.uk

**Abstract**
Self-consistent transport simulation of ITER scenarios is a very important tool for the exploration of the operational space and for scenario optimisation. It also provides an assessment of the compatibility of developed scenarios (which include fast transient events) with machine constraints, in particular with the poloidal field (PF) coil system, heating and current drive (H&CD), fuelling and particle and energy exhaust systems. This paper discusses results of predictive modelling of all reference ITER scenarios and variants using two suite of linked transport and equilibrium codes. The first suite consisting of the 1.5D core/2D SOL code JINTRAC [1] and the free boundary equilibrium evolution code CREATE-NL [2,3], was mainly used to simulate the inductive D-T reference Scenario-2 with fusion gain Q=10 and its variants in H, D and He (including ITER scenarios with reduced current and toroidal field). The second suite of codes was used mainly for the modelling of hybrid and steady state ITER scenarios. It combines the 1.5D core transport code CRONOS [4] and the free boundary equilibrium evolution code DINA-CH [5].

## 1. Introduction

Self-consistent predictive simulation of ITER scenarios is a very important and widely used tool for the exploration of the operational space and for scenario optimisation [5-17]. It can also provide an assessment of the compatibility of the developed scenarios (including fast transient events) with machine constraints; in particular with the poloidal field (PF) coil system, with heating and current drive (H&CD), fuelling, particle and energy exhaust systems. Previous results were centred either on predictive transport modelling of ITER scenarios or on the prediction of equilibrium evolution and plasma control. However, the quality and trustworthiness of the prediction of the plasma and plasmas systems behaviour can be greatly improved when the best combination of high quality 1.5D core transport codes, which employ complex theory-based transport models, are combined with state of the art free boundary equilibrium codes. This paper discusses results of such "combined" predictive modelling of all reference ITER scenarios and variants using two suites of linked transport and equilibrium codes. The first suite consisting of the 1.5D core/2D SOL code JINTRAC [1] and the free boundary equilibrium evolution code CREATE-NL [2,3], was mainly used to simulate the reference 15 MA inductive D-T burn scenario with fusion gain Q=10 and its variants in H, D and He (including ITER scenarios with reduced current and toroidal field). The second suite of codes was used mainly for the modelling of hybrid and steady state ITER scenarios. It combines the 1.5D core transport code CRONOS [4] and the free boundary equilibrium evolution code DINA-CH [5].

The rest of the paper is organised as follows. Chapter 2 is devoted to a detailed description of the reference 15MA baseline inductive burn scenario and its variants. Chapter 3 summarises main assumptions and models used in predictive transport modelling of these scenarios both in transport code JINTRAC and in free boundary code CREATE-NL. Chapter 4 summarises the main results of predictive modelling of these scenarios highlighting new findings and remaining issues. Chapter 5 describes the main features of Hybrid and Steady State (SS) as well as main assumptions and models used to simulate these scenarios by a combination of core transport code CRONOS and free boundary code DINA-CH. Chapter 6 summarises the main results of the modelling of Hybrid and SS scenarios. Finally, Chapter 7 briefly summarises general results, discusses remaining issues and suggests future activities.

**2. Specification of the reference 15MA baseline inductive burn scenario and its variants.**

The reference 15 MA inductive burn scenario with $Q_{fus}$=10 is considered as the reference ITER baseline scenario. Its main characteristics are summarised in tables 1-2 and we will refer to it as to a Case#001. During the ramp-up phase, an early transition from the limited to the diverted phase is sought, applying auxiliary heating right afterwards to reduce resistive flux losses. When a current maximum of 15 MA is reached, the auxiliary heating is increased to a maximum level to assure a quick transition to a good quality type-I ELMy H-mode regime. As soon as the plasma density has reached a target level close the Greenwald density limit and the fusion process has fully developed, the auxiliary heating is reduced to reach the ITER target of getting close to $Q_{fus}$ ~ 10. The flat-top phase is maintained until the PF coil flux charge limit for a safe H-L transition and plasma ramp-down is reached. The current and plasma density are then ramped down and auxiliary heating is gradually decreased. The H-L transition is assumed to occur at the beginning of ramp-down when the plasma energy content is still at its maximum level in order to get close to the worst possible transition that can happen during the discharge. The main interest here is to test whether the plasma position control can be maintained by the PF coil control system also under extreme conditions. Like for current ramp-up, the transition between the diverted and limited plasma at current ramp-down takes place at a low current level, with the aim to increase the flexibility of the auxiliary heating systems for plasma control avoiding the lower heat limit for the plasma-facing components in limiter configuration. The plasma discharge is simulated until the end of current ramp-down when high $q_{95}$>10 regimes are reached, for which the transport model under consideration may not be applicable.

The Case#002 scenario only differs from Case#001 for the current ramp-down phase. The intention for this ITER baseline variant is to ramp down $I_{pl}$ more slowly. This could be beneficial for fuelling, as the Greenwald density $n_{GW}$ scales with $I_{pl}$ and it is not sure if the plasma particle content can be depleted quickly enough at high $|dI_{pl}/dt|$ for the plasma density to stay below the density limit $n_e<n_{GW}$. Another advantage is the evolution of inductive current. Current density profiles are less peaked at smaller $|dI_{pl}/dt|$ during ramp-down.

The Case#003 scenario is characterised by very short current ramp-up and ramp-down phases at maximum ramp rates that are achievable without violating PF coil voltage limits. The corresponding ramp-up/-down duration $\tau_{ramp}$ was found to be $\tau_{ramp} \approx$50 - 60 s.

The Cases#1-3 assume that transition from limiter to divertor configuration during current ramp happens at $I_{pl} \approx$ 4 MA. To examine the sensitivity of plasma properties on the $I_{pl}$ level when the transition between the limited and diverted phase takes place, another variant typed Case#004 with transition to/from diverted phase at plasma ramp-up/-down foreseen at $I_{pl} \approx$ 7 MA has been developed. Again, special attention was paid to the flux consumption balance. Also, in the limiter phase the power flux crossing the separatrix was kept below the limit to avoid damage to the first wall. Simulations have been also performed for the ITER pre-activation phase. The following three scenarios have been modelled:
- Case#005: H plasma, $I_{pl}$ = 15 MA at flat-top, $B_0$ = 5.3 T.
- Case#006: H plasma, $I_{pl}$ = 7.5 MA at flat-top, $B_0$ = 2.65 T.
- Case#007: He plasma, $I_{pl}$ = 7.5 MA at flat-top, $B_0$ = 2.65 T.

with the emphasis to assess if high quality type-I ELMy H-mode conditions can be achieved during initial, pre-activation phase of ITER operation.

Based on simulations for the 15 MA ELMy H-mode ITER baseline scenario (Case#001) and sensitivity studies that have been made to find potentials of optimisation with respect to either neutron yield or the fusion energy production per discharge $W_{fus}$, new scenarios referred to as Case#008-010 have been developed where all actuators that have been identified to allow an enhancement in $W_{fus}$ have been combined, trying to minimise disadvantageous side effects. $W_{fus}$ can either be maximised by increasing the fusion reaction rate in the burning phase by increasing plasma current to $I_p$=17MA or by extending the pulse duration by optimising current ramp down rate. Differences between Cases#008-010 can be summarised as follows:
- Case#008: $I_{pl}$ = 15 MA at flat-top, $V_{loop}$ = 0.0 V for $I_{pl}$ = 15→5 MA at ramp-down
- Case#009: $I_{pl}$ = 17 MA at flat-top, $V_{loop}$ = 0.0 V for $I_{pl}$ = 17→5 MA at ramp-down
- Case#010: $I_{pl}$ = 17 MA at flat-top, $V_{loop}$ = 0.0 V for $I_{pl}$ = 13→5 MA at ramp-down

Predictive modelling of the reference 15MA inductive burn scenario and its variants was also supplemented by an extensive analysis and modelling of fast transient phenomena, which include L-H and H-L transitions, strong ELMs, triggered by a sawtooth crash and minor disruption. This study allowed us to test the ability of presently foreseen ITER position and vertical stability control system to cope with fast variation in some major plasma parameters such as total stored energy and plasma inductance.

## 3. Main assumptions and models used in predictive transport modelling of 15MA inductive burn scenario and its variants

The JET integrated transport code suite JINTRAC has been used to model the ITER baseline scenario in combination with the free boundary equilibrium code CREATE-NL. CREATE-NL/JINTRAC coupling can be used in a strong or weak form. In the "strong" form, coupling is made time step by time step. This form of coupling requires high computational times and can be reasonably run for a limited number of time steps, i.e. full scenario with a time step of 100-200 ms (no eddy currents) or short time windows with a time step of 1-2ms (including eddy currents and vertical stability controller (VS) simulation). The need to simulate volt-second (Vs) consumption on significant time windows, forced us to also develop the so called "weak" coupling between JINTRAC and CREATE-NL. In particular JINTRAC has been run on the basis of a certain sequence of shapes and plasma currents provided by CREATE team. Then a free boundary simulation has been re-run with CREATE-NL using kinetic profiles from JINTRAC. The procedure was closed verifying that JINTRAC results are insensitive enough to shape variations simulated with the free boundary closed loop CREATE-NL runs. In both coupling modes, CREATE-NL provides equilibrium-related information to JINTRAC, which passes profile data about the plasma state back to the equilibrium code. Plasma shape control is simulated by CREATE-NL by calculation of PF coil correction currents. These are put on top of nominal currents that are pre-calculated on basis of the desired scenario shape evolution. In JINTRAC, the evolution of core plasma conditions is simulated in a semi-predictive way, with transport equations being solved for current diffusion and heat transport but the evolution of the shape and the volume-average level of main ion and impurity densities being prescribed. To make sure that the assumed density level and profile shape are reasonable and compatible with the ITER fuelling system, complementary fully predictive simulations have been carried out giving satisfactory agreement with the semi-predictive calculations. The Bohm/gyroBohm (BgB) model without non-local multiplier is used for the prediction of plasma transport in L-mode [18]. This choice is motivated by extensive model validation and benchmarking efforts that have been carried out within the ITER Scenario Modelling Working Group during the past few years (see e.g. [19,20]). Unlike alternative transport models, the BgB model has turned out to predict L-mode plasma confinement for a wide variety of tokamak machines and discharge configurations reasonably well. For the simulation of H-mode phases, the GLF23 model was selected, as it is the best validated theory-based transport model for stiff plasma regimes [21]. Transition to H-mode is triggered when the total net plasma heating exceeds L-H transition power threshold from Martin et al. [22]. The GLF23 transport is assumed to be almost fully suppressed within edge transport barrier (ETB) after L-H transition. The ETB width was fixed to a level that allows fulfilling the condition $Q_{fus} \sim 10$ during flat-top for Case #001 and is broadly in line with what is prescribed by EPED model [23]. It was kept the same in all further simulations. A simple empirical transport model is used for the ETB zone to describe ELM-induced transport enhancement within ETB. Namely we enhance all transport coefficients within the edge barrier when normalised edge pressure gradient $\alpha$ approaches ballooning stability limit $\alpha_c$ inferred from MHD stability analysis. We then average this transport enhancement in time and adjust its level to ensure that plasma stays close to a prescribed critical normalised pressure gradient $\alpha_c$ (we call this approach a "continuous ELM" model). Different approach was used when we simulated large, "uncontrolled" ELMs. In this case transport enhancement was selected to be very strong and very short-lived. Heat deposition profiles were either calculated by JINTRAC-internal models or prescribed from external sources [24]. The influence of fast particles is taken into consideration. In particular, the alpha thermalisation process, which can delay the energy decay and inwards plasma movement after the H-L transition, is accounted for in the simulations. Sawtooth reconnection is modelled by application of the Kadomtsev model. Flux consumption is determined following the axial representation described in [25] with a splitting in contributions from inductive flux, resistive on-axis losses and sawtooth-induced losses.

One of the novel elements of this modelling was predictive simulation of fast transient phenomena, such as L-H and H-L transitions as well as predictive modelling of D and T densities and He ash accumulation. Whereas the transition from H-mode to L-mode is prescribed in Cases#001-010, a self-consistent L-H and H-L transition model with the power threshold $P_{L-H}$ scaling from Martin et al. [22] is used in supplementary studies. Depending on the assumptions for the heat flux $P_{net}$ to be compared with $P_{L-H}$ (which, in some cases, include intermediate transition from L to type-III ELMy H-mode with a lower level of a critical normalised pressure gradient $\alpha_c$), the behaviour of the transition can change featuring either a slow or a fast change in energy content and some dithering between different confinement states with varying duration and frequency. This dithering is caused by the increase in $P_{net}$ due to heat pulse propagation after an H-L transition and the reduction in heat flux when the heat input to the plasma is spent on increasing the plasma energy in the core after an L-H transition and a formation of high quality edge transport barrier (ETB).

All the free boundary simulations were produced with the CREATE-NL code [2, 3]. This is an axisymmetric free boundary equilibrium code based on the numerical solution of Grad-Shafranov (GS) equation. It implements first order finite elements method based on a Newton method for the solution of the free boundary nonlinear associated problem. A numerical calculation of the Jacobian matrices is adopted. CREATE-NL can account for the presence of eddy currents and iron. In standard running mode it accepts arbitrary current density profiles specified by means of ff' (f = RBΦ/μ0) and p' (p being plasma pressure) profiles in the GS approximation coming from JINTRAC. As for the design of the controllers and optimization of the nominal currents to drive plasma through the scenarios, CREATE adopted several routines developed in the Matlab/Simulink environment which helped during the iterative process required to design control. Namely, the feedback control action used for the simulations is composed of three parts:
- A vertical stabilization controller (VSC) controlling plasma vertical position or speed using the VS3 circuit voltage (see [6-8] for a description of ITER poloidal fiend system (PF system);
- A current control (CC) controlling currents in the active coils (central solenoid (CS) coils and P1-P6 coils) using voltage on active coil power supplies;
- A shape controller (SC) controlling gaps between separatrix and the wall or other shape parameters using currents (reference currents to the CC) on the active coils.
The controllers have been designed and used under the following hypotheses:
- A feed forward action in current is available, i.e. there is an *a-priori* approximated knowledge of the nominal sequence of control currents driving the plasma through the scenario. The feed forward control action is a piecewise linear function of time;
- Full availability of plasma quantities including gaps (ideal reconstruction of shape);
- No noise on measurements,
- Ideal plasma current control (assigned Ipl time history during the simulation),
- Simplified dynamics of the vertical stability (VS) converters. The controllers developed are designed for the purpose of controlling the plasma in simulations and are not meant to be implemented, as they are, on a real tokamak.

It is worth mentioning here that since the information between JINTRAC and CREATE-NL is exchanged discretely in time, caution should be exercised in order to preserve integrity of time evolution of plasma parameters in evolving plasma shape. This is easily achieved in slowly varying plasma when plasma pressure evolves faster than current profile. JINTRAC uses numerical scheme, which preserves q-profile during equilibrium recalculation and evolve it in accordance with Faraday equation between successive recalculations of equilibrium [1]. Fast transient (ELMs, L-H and H-L transition are more difficult because edge current evolves on a time scale comparable to transport time scale. Analysis shows however that conservation lows can be still preserved during such fast transients by using more frequent recalculation of equilibrium.

**4. The main results of predictive modelling of 15MA inductive burn scenario and its variants.**

O*ptimisation of the plasma current ramp-up and ramp-down and sensitivity to the plasma current ramp rate:* A sensitivity scan for the ramp-up phase was performed with respect to $dI_{pl}/dt$, the amount of auxiliary heating, the density level and boundary conditions at the separatrix. For higher ramp rates and off-axis heating power, lower values for the internal inductance li(3) and a reduction in Vs consumption can be achieved (see figure 1). Plasma profiles at the end of ramp-up are insensitive to assumptions for boundary conditions and not very sensitive to assumption for fuelling. If the current is ramped up at the maximum rate that can be provided by the PF coil system, the flat-top performance is degraded compared to the reference scenario. This degradation is caused by the decrease of the ratio s/q between shear and safety factor in the early phase of flat-top, causing an amplification of micro turbulences in the core according to experimental observations [26] and predictions with GLF-23 as detailed in [27,28] and a reduction in fusion performance by up to ≈60%. The same reason causes initial degradation of the flat top performance in case of early L-H transition (see

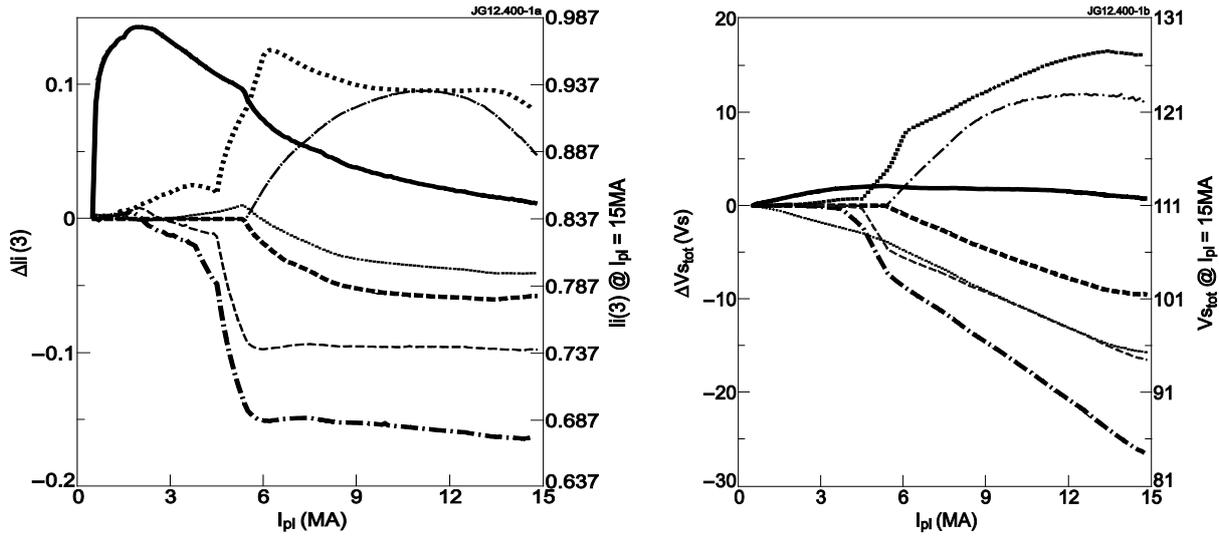

*Figure 1 - deviation in li(3) (a) and total Vs consumption (b) during current ramp-up with respect to the reference case, for 20% of $n_{GW}$ (thin dotted), 0 MW (thin chain) and 20 MW of ECRH (thick dash), 30 eV of initial boundary temperature (thick solid line, reference case: 100 eV), -30% (thick dots), +30% (thin dashed) and +66% of $dI_{pl}/dt$ (thick chain).*

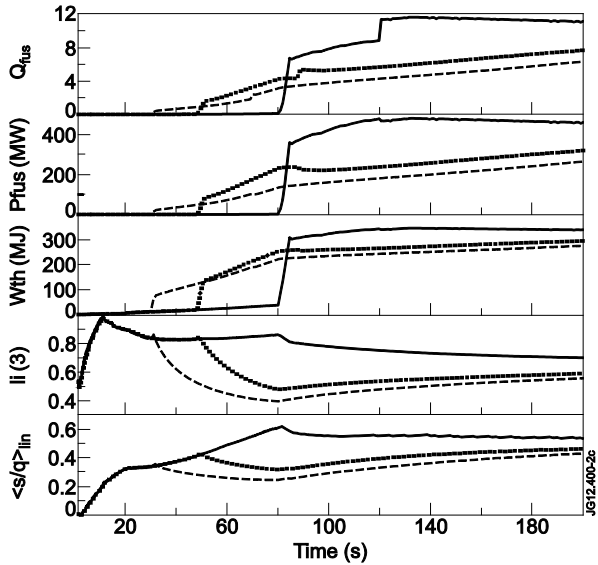

*Figure 2 - from top to bottom: fusion Q, $P_{fus}$, $W_{th}$, li(3) and line-averaged s/q ratio for the reference case with L-H transition at 15 MA (solid), and two simulation cases with early L-H transition at 10 MA (dotted) and 7 MA (dash). In the latter case a Greenwald density fraction of 60% was applied to avoid the NB shine-through limit after the transition.*

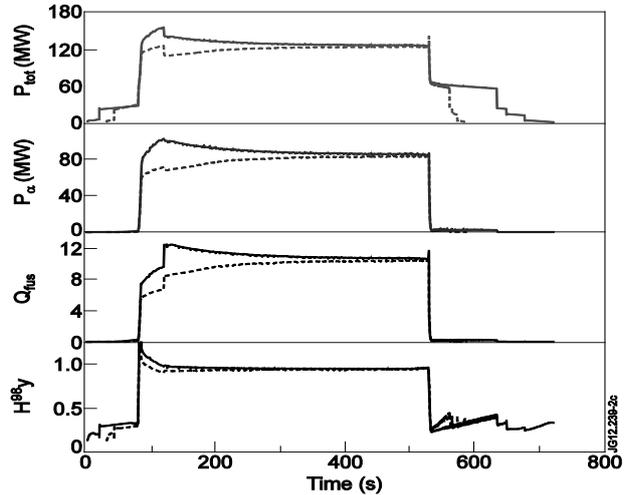

*Figure 3 - total deposited heating power, alpha heat deposition, fusion gain Q and H98y (from top to bottom) for the complete reference scenario (solid) and the maximum ramp-rate scenario (dashed).*

Figure 2). In the current ramp-down, the application of the maximum available current ramp rate helps to improve the improve the Vs balance, but plasma control related to vertical stability and density pump-out becomes more demanding (see figures 3-4). Some of these conclusions have been confirmed in the experiment [29]. The effect of a variation in ramp rate on li(3) and Vs for the current ramp-down phase is shown in figure 5. Resistive current losses are dominant in the late phase of ramp-down due to the decrease in electron temperature. With a late divertor-limiter phase transition, high $P_{AUX}$ can be maintained at lower $I_{pl}$ which helps to shape the current profile and reduce resistive losses. Sawtooth-induced Vs consumption plays a role in the early phase after the H-L transition, as q scales inversely with plasma current.

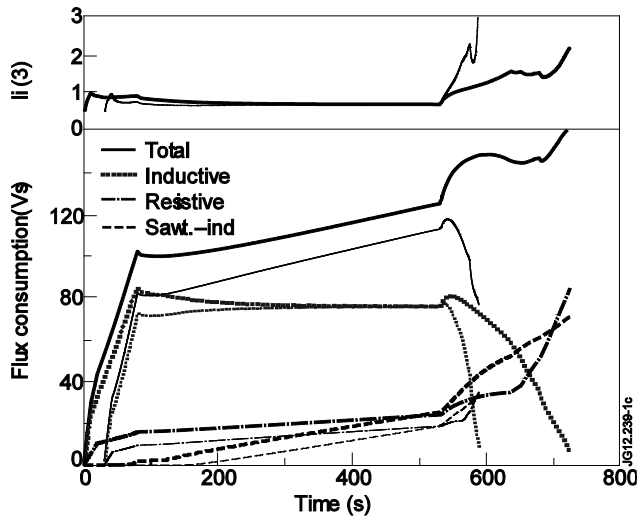

*Figure 4 - li(3), total, inductive, resistive and sawtooth-induced Vs consumption (from top to bottom) for the complete reference scenario (thick lines) and the maximum ramp-rate scenario (thin lines).*

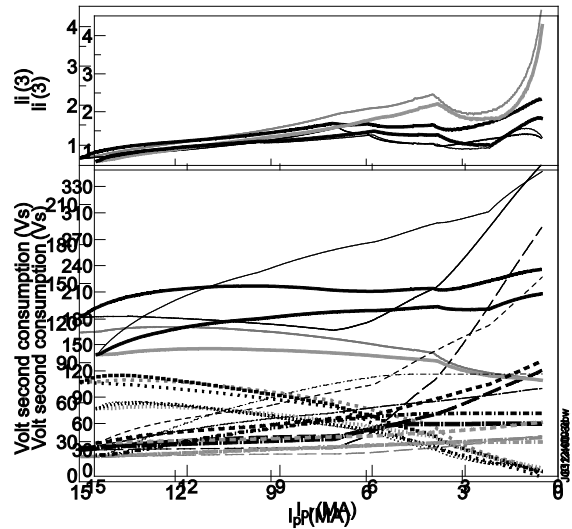

*Figure 5 - top: li(3), bottom: total (solid), inductive (dotted), resistive (dashed) and sawtooth-induced (dash-dotted) Vs consumption, for slow (thin lines), medium (thick black lines, reference scenario) and fast ramp-down (thick grey lines). The ramp-down periods are 400 s, 200 s and 60 s resp..*

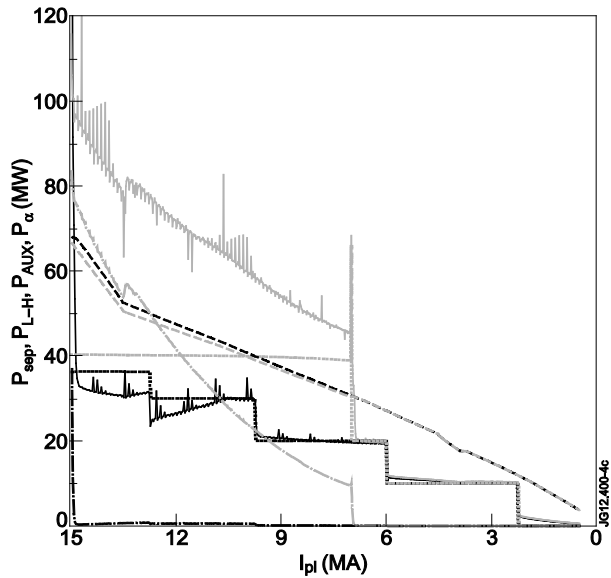

*Figure 6. Power crossing the separatrix (solid), external heating (dotted), L-H transition power threshold [22] (dashed) and alpha heating power (chain), with L-H transition at 15 MA (black) and 7 MA (grey).*

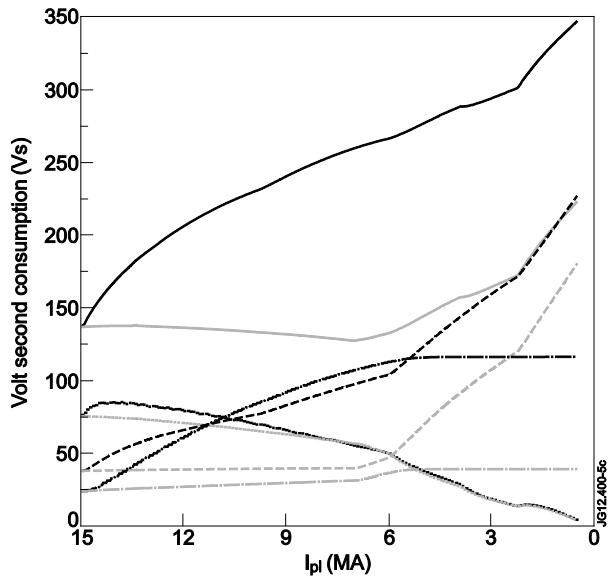

*Figure 7 - total (solid), inductive (dotted), resistive (dashed) and sawtooth-induced (chain) Vs consumption for slow ramp-down (ramp-down period: 400 s) with H-L transition at 15 MA (black) and at 7 MA (grey).*

*Sensitivity to timing of the L-H transition:* An anticipated early transition to H-mode during current ramp-up helps to reduce Vs consumption. It may also reduce the risk of low-frequency NTM-triggering sawteeth by reduction of the radial position of $q = 1$ [30] and facilitate plasma control after the transition due to reduced alpha heating at lower currents. However, as mentioned before, early L-H transition leads to a temporary reduction in plasma performance during initial phase of current flat top (see Figures 2 and 3). Further analysis shows however (see Optimised Scenarios on pp 12-13) that temporary loss of fusion power vanishes after current redistribution time. Moreover, since using early L-H transition allows significant reduction in Vs consumption, initial losses in neutron production is fully compensated by the lengthening of the burn phase.

*Sensitivity to timing of the H-L transition:* In order to reduce resistive and sawtooth-induced Vs losses, it is highly preferable to maintain H-mode conditions during the current ramp-down for as long as possible. The current ramp-down period could then (and only then) be considerably extended to more than 400 s. With the help of alpha particle heating in the early phase of ramp-down and strong auxiliary heating later on, it appears to be possible to delay the H-L transition until a current level drops below 7 MA ($\rightarrow$ figures 6-7). This result was confirmed in semi-predictive simulations for density levels in a range of 40-80% of $n_G$.

*Self-consistent simulations of the plasma evolution after the L-H / H-L transition and sensitivity to H-L transition dynamic:* One of the critical points observed when simulating 15 MA Scenarios (Case#001 or Case#003) was the H to L transition. In facts voltage saturation during a 15 MA H-L transition might result in a plasma wall contact or at least can bring the plasma very close to the first wall. A crucial role during this phase is played by the beta-poloidal drop amplitude and its decay rate, by the maximum allowed voltage on active coils, and by the VS and shape control law adopted. As for the shape control law a feedback plus feed forward control law was used for all the proposed simulations due to the benefits that this kind of strategy can provide in managing the discharge also in the presence of constraints and limitations. A scan in the foreseeable thermal energy decrease rate $dW_{th}/dt$ after the H-L transition, depending on possible transitional plasma confinement states like an intermediate type-III ELM regime, the speed of the reduction in fusion power $P_{fus}$, the alpha particle thermalisation and energy confinement times and the level of $P_{AUX}$ reduction, was carried out and the most extreme cases with the maximum conceivable $dW_{th}/dt$ have been determined.

Seven cases have been selected for further analysis with CREATE-NL. First of all, two reference cases: one for the attenuated H-L transition in Case#001 with a long transitional period (henceforward referred to as case HL#1) and for the most extreme conceivable, fastest H-L transition (HL#2) at the end of the flat-top baseline scenario, were re-run. For the better identification of distinct causes for control problems and the weight of their influence during the transition, these two cases have been rerun at time-invariant density, shape, and plasma current at an earlier stage of the flat-top phase where the transformer is not yet operating close to the available flux limit (HL#3 resp. HL#4). Finally, three additional H-L transition cases with an attenuated (HL#5) and a sharp (HL#6) drop in energy and another case where intermittent transitions take place at a higher energy level in H-mode (HL#7) have been calculated using a self-consistent H-L transition model. In cases HL#1-6, 33 MW of NBI + 7 MW from ICRH are applied at flat-top, whereas total RF power is increased to 30 MW (20MW of ICRH plus 10 MW ECRH) for HL#7 before the transition. In case of the fastest possible transition (HL#2, HL#4, and HL#6) as well as in HL#5 and HL#7, auxiliary heating is completely switched off when the plasma returns to L-mode. In HL#1 and HL#3, 33 MW of NBI + 20 MW ICRH are applied after the transition with the intention to reduce resistive flux consumption.

GLF-23 in combination with the continuous ELM model is used for H-mode, and the Bohm/gyroBohm model without non-local multiplier for L-mode. Whereas the transition from H-mode to L-mode is prescribed in cases HL#1-4, a self-consistent H-L transition model with the $P_{L-H}$ scaling from Martin et al. is used for HL#5-7. Different assumptions can be used to trigger H-L transition. If we assume that H-L transition is triggered when the heat flux through the separatrix drops below $P_{LH}$ the transition features relatively slow reduction in energy content and some dithering between different confinement states with varying duration and frequency (case HL#5). If the transition is triggered when the heat flux deeper inside plasma drops below $P_{LH}$, then transition can be very fast (case HL#6). Depending on the assumptions for the position of the heat flux to be compared with $P_{L-H}$, the behaviour of the transition can also exhibit some spontaneous temporary back-transitions to H-mode (case HL#7).

Some simulation results from JINTRAC for the cases HL#1-7 are shown in Figs. 8-9. Depending on the heating conditions after the H-L transition, $\Delta\beta_{pol} \approx 0.6$-0.7, $\Delta li(3) \approx 0.20$-0.25, and $\Delta W_{th} \approx 270$-300 MJ for HL#1-4, and $\Delta\beta_{pol} \approx 0.6$-0.8, $\Delta li(3) \approx 0.10$-0.25, and $\Delta W_{th} \approx 290$-370 MJ for HL#5-7. For the "extreme" transition cases the exponential $\beta_{pol}$ decay time varies between $\approx 0.40$ s (HL#2, HL#4) and 0.65 s (HL#6). CREATE-NL simulations revealed (see below) that the change and the rate of change in $\beta_{pol}$ are in a range where plasma shape control can become a challenge. The plasma column quickly approaches the inner wall after the transition, requiring a strong action from the central solenoid to decelerate its radial inward movement for the prevention of its contact with the inner wall. If the outward driving force due to the feedback action of the PF coils is too strong though, one might risk a contact with the outer wall just a few seconds later. The PF coil current feedback control action therefore needs to be carefully designed.

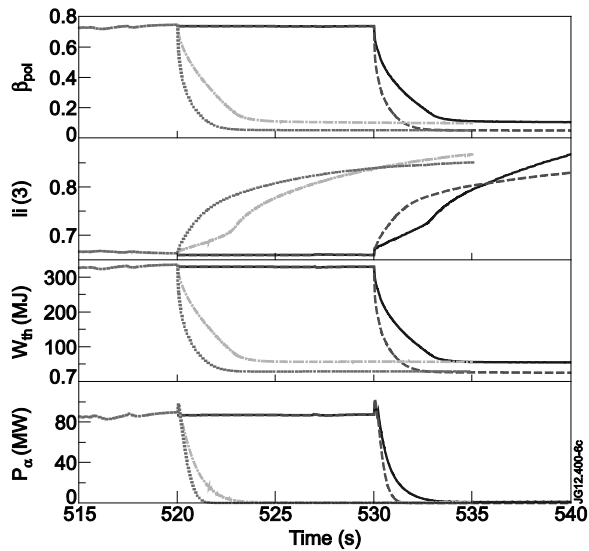
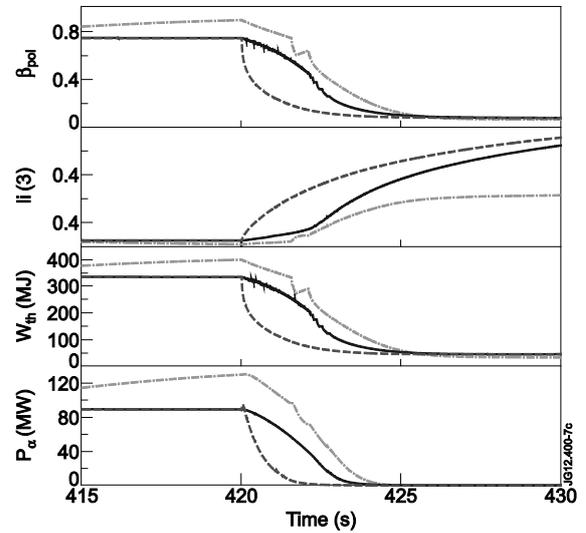

*Figure 8 – From top to bottom: $\beta_{pol}$, li(3), $W_{th}$ (in MJ), and $P_\alpha$ (in 100 MW), for the H-L transition cases HL#1 (solid), HL#2 (dash), HL#3 (chain), HL#4 (dotted); output for HL#3-4 is shifted in time.*

*Figure 9 – From top to bottom: $\beta_{pol}$, li(3), $W_{th}$ (in MJ), and $P_\alpha$ (in 100 MW), for the H-L transition cases HL#5 (solid), HL#6 (dash), and HL#7 (chain).*

The controllability analysis was primarily made with reference to the plasma-wall distance at the equatorial plane that is the most critical shape parameter during HL transitions. Since the multivariable controller used in this study allows the simultaneous control of many gaps, the control action was tuned so as to provide good performance on the radial inner gap relaxing requirements on other important shape control parameters as strike points which however do not seem to be a problem in the proposed simulations. The following degree of freedom in the control law that can have an impact on its efficiency and effectiveness were assumed in CREATE-NL analysis:
- Time update of the current nominal feed-forward control action (1s, 5s, mixed strategies).
- Possible delays in applying the feed-forward control action.
- Possible anticipation of the feed-forward control action in time.
- Scheduling of the feed-forward control action with beta-poloidal estimate.

All the proposed sensitivity studies are concerned with the balance between feed forward and feedback actions. As for the feed forward current update sampling period we consider τ = 1s as the shortest period for the current update. In fact going under this period would require very precise a-priori knowledge of the plasma behaviour along the scenario and of the eddy currents evolution, which can be hardly achieved. The absence of updates of the nominal pre-programmed control currents requires minor corrections, in the order of modelling uncertainties, that can be generated by the feedback control action. In addition the current control is designed with time constants in the order of tenth of seconds to avoid saturations of the PF coils power supply voltages. With τ = 1s we register the best performance during the initial phase of the beta-poloidal drop. A side effect of updating at this (relatively) high frequency is that, if we have a proportional integral action on the gap control loop, after an initial phase when the feed forward and feedback control actions produce a joined effort to push the plasma away from the internal wall, there is a residual feedback effect that causes the plasma to move towards the external wall. This effect could be counteracted in different ways:

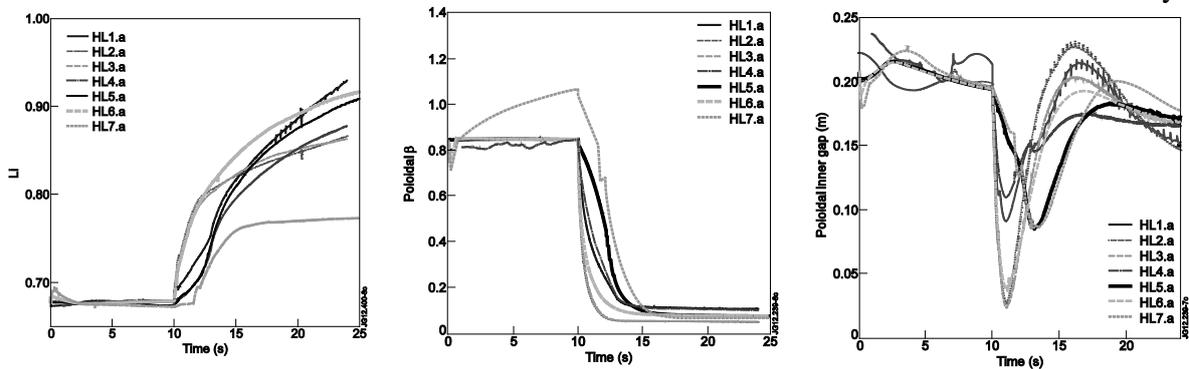

*Figure 10. Time evolution of internal inductance li(3) (left), beta poloidal (centre) and radial inner gap (right) for the cases HL#1-7)*

- A lower frequency update of the nominal currents causes a weaker "push away from the internal wall" effect;
- A different choice of the feedback control law can help. There are two possibilities: to make a different tuning of the proportional-integral action, or assume a different strategy in which voltage control is used for fast transients.

If we update the feed forward with τ = 5s the main risk we run is that, during transients, the feedback action (as tuned for the reference controller used in this study) is not able to compensate the disturbance action. A better strategy for the feed forward update may be to have a 1s update just after the transition starts, and then have a 9 seconds update to avoid that feed forward plus feedback can move the plasma towards the external chamber wall. The effect of a delay on the application of the feed forward currents in correspondence of the H-L transition starting point is dangerous for fast transients. On the other hand the possibility of anticipating the effort needed to counteract the plasma movement toward the inner wall in correspondence of beta drop does not seem to give significant advantages. The possibility to apply the feed forward control action as a function of beta poloidal has also been studied. This approach requires an online estimate of beta poloidal which can be assumed available in real world with some small delay and approximation. This approach seems to be promising with respect to performance and gives n improved capability to react to unpredicted situation such as unexpected or aborted transitions. It is worth to notice that for all the proposed cases there is a slight exceeding of P6, P5 vertical and/or CS separation forces. This issue has been resolved by better optimizing the nominal pre-programmed current.

Figure 10 shows an example of CREATE-NL simulation for the cases HL1-HL7 with τ=1s for feed forward current update after H-L transition. One can see that plasma avoids wall contact even in the most dangerous, fastest transitions, although the minimum distance between separatrix and the inner wall temporarily drops below 7 cm, which is considered as the minimum safe distance. It is worth to notice that the proposed controller is still far to be optimized and the results obtained depend both from the control action and the used plasma model However the proposed methodology can be refined with further studies on numerical models and experimental ITER data as they will be available

Another example of supplementary activity includes fully predictive simulations (i. e. solving the transport equations both for temperature and density) of the reference ITER baseline H-mode scenario and for the Hybrid scenario. In the simulations particular attention was paid to whether the prescribed density evolution during the density ramp-up and the target value during the burn phase could be achieved by means of pellet injection in combination with a feed-back mechanism on the plasma average density.

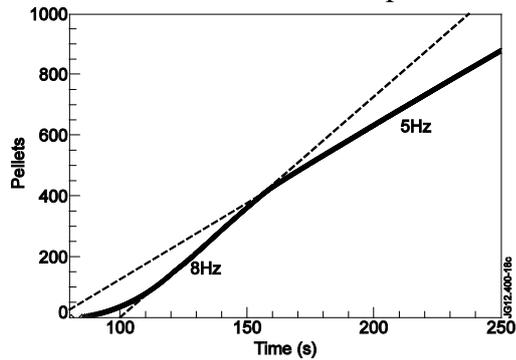

*Figure 11 Pellet injection rate as function of time needed to increase plasma density to a target value in inductive burn scenario.*

The transport model adopted in these simulations was the semi-empirical Bohm/gyro-Bohm transport model [18] where the coefficients in front of the Bohm and gyro-Bohm part of the thermal diffusivities (originally calibrated against JET discharges) where adapted to recover the target parameters of the ITER scenario under consideration and an anomalous pinch velocity, of the form $v_{inw}=0.5\ D\ r/a^2$, was introduced to reproduce the density peaking predicted by physics based models like GLF23. In particular, for the base line H-mode scenario, we multiplied the Bohm contribution to the electron and ion thermal conductivity a factor of 4 and the gyro-Bohm contribution by a factor of 8 and 16 respectively (for details see [31]), whereas, for the hybrid scenario, in order to recover a confinement enhancement factor $H_{98}$ of ~1.3, we divided by a factor of 4 the gyro-Bohm part of the ion and electron thermal diffusivities.

The simulation of the density ramp-up shows that a variable pellet injection frequency is necessary to obtain a gradual density increase (see Figure 11). It can be seen that, to achieve a density ramp-up similar to the one prescribed in the semi-predictive simulations described in this paper, one has to increase the pellet frequency up to a maximum of ~8 Hz and then reduce it to a steady-state value of ~5 Hz.

The simulations of the flat-top of the base line H-mode scenario show that typical ITER fuelling pellets (mass: $7.5\ 10^{21}$ particle, composition: 50/50 DT and deposition profile centred around ρ=0.95) should guarantee the sustainment of the target density prescribed in the semi-predictive simulations. The injection frequency could vary between 1.5 Hz and 4 Hz depending on the model adopted for the particle transport during ELMs.

Similar simulations for the flat-top of the hybrid scenario indicate that the pellet frequency required to achieve the target density prescribed in the semi-predictive simulations could be as high as 6 Hz.

*Optimisation of baseline inductive burn scenario.* Based on simulations for the 15 MA ELMy H-mode ITER baseline scenario (which was optimised to achieve maximum fusion gain for a specified time interval) and sensitivity studies that have been made to find potentials of alternative optimisation with respect to the fusion energy production per discharge $W_{fus}$, new scenarios have been developed where all actuators that have been identified to allow an enhancement in $W_{fus}$ have been combined, trying to minimise disadvantageous side effects.

$W_{fus}$ can be maximised either by increasing the fusion reaction rate in the burning phase or by extension of the pulse duration. A summary of techniques for the maximisation of $W_{fus}$ is given below. Optimisation techniques are not fully exploited if there is a risk that they could lead to the violation of operational constraints (related to heating, fuelling, MHD stability, confinement conditions, PF coil current control etc.).

*Simulation conditions.*
*Current ramp-up*: Earliest possible transition to a diverted plasma configuration (at $I_{pl} = 4$ MA), current ramp at the maximum supportable rate $dI_{pl}/dt \approx 0.3$ MA/s, low fuelling: $n_{e\,lin.\,avg.} \approx 0.25 \cdot n_{GW}$, application of broad on-axis ECRH with additional 16.5 MW of NB power as soon as the density rises above the NBI shine through limit, keeping $P_{AUX}$ just below $P_{L-H}$, L-H transition at the end of current ramp-up.

*Flat-top*: Low fuelling: $n_{e\,lin.\,avg.} \approx 0.65 \cdot n_{GW}$, application of full available auxiliary heating power, assumed to consist of 20 MW of ECRH, 20 MW of ICRH (both deposited in the central plasma region), and 33 MW of NBI.

*Ramp-down*: $V_{loop} = 0.0$ V or close to zero until $I_{pl} \approx 5$ MA, prescribed $dI_{pl}/dt$ for $I_{pl} < 5$ MA, maintenance of H-mode until $I_{pl} \approx 7.5$-8.5 MA, same fuelling assumptions as for current flat-top: $n_{e\,lin.\,avg.} \approx 0.65 \cdot n_{GW}$, gradual decrease in $P_{AUX}$, late transition to limited configuration at $I_{pl} = 4$ MA. (see Fig. 12)

Three optimised scenarios Cases ##008-010 have been tested and compared to the ITER baseline scenario (Case #001). Differences in the scenario configuration can be summarised as follows:

- Case #008: $I_{pl} = 15$ MA at flat-top, $V_{loop} = 0.0$ V for $I_{pl} = 15 \rightarrow 5$ MA at ramp-down
- Case #009: $I_{pl} = 17$ MA at flat-top, $V_{loop} = 0.0$ V for $I_{pl} = 17 \rightarrow 5$ MA at ramp-down
- Case #010: $I_{pl} = 17$ MA at flat-top, $V_{loop} > 0$ for 17MA$>I_{pl}>$13MA; $V_{loop} = 0.0$ V for $I_{pl} = 13 \rightarrow 5$ MA at ramp-down

All three scenarios are not yet fully optimised in the ramp-down phase, where a constant $V_{loop}$ close to 0.0 V has just been applied for reasons of simplicity. The actual flux limits are estimated to allow an additional extension of the ramp-down phase by several hundreds of seconds.

Simulation results are shown in Figs.12-15. It seems to be advantageous to operate at the maximum flat-top current for which a safe and stable operation can be assured, as not only fusion power gets considerably enhanced as expected, but also resistive and sawtooth-induced flux consumption $\Psi_{res}$ and $\Psi_{saw}$ do not increase substantially in the flat-top phase. For a given number of Vs that can be saved, the amount of time by which the flat-top phase could be extended is almost the same for flat-top currents of 15 and 17 MA. Expressed in numbers, one Vs is consumed in quasi-steady state flat-top conditions within $\approx$ 23-25 s for all optimised scenarios (Cases ##008-010). Despite the very similar flux consumption properties, the flat-top duration $\tau_{flat-top}$ needs to be shortened at higher currents to stay within the safe limit for the PF system. Comparing scenarios with $I_{pl} = 17$ MA and $I_{pl} = 15$ MA with similar flux limit assumptions, we conclude that the associated reduction in fusion power in 15MA Case #008 would overweigh the gain in $\Delta t(I_{pl}>15$ MA), and that the $I_{pl}(t)$ evolution in Case #010 at $I_{pl} > 15$ MA must be close to the optimum one. As discussed earlier, the flux optimising measures at current ramp-up have the side effect of flattening the current density profile. This leads to a lower s/q in the outer plasma region and degradation in confinement in accordance with transport theory predictions and experimental observations. Nevertheless, it seems to be reasonable to disregard the tailoring of the q profile and instead minimise $\Psi_{res}$ and $\Psi_{saw}$ at current ramp-up stage. Comparing Case #008 with Case #001, $\Psi_{res}+\Psi_{saw}$ drops from 17.6 Vs down to 7.4 Vs at the beginning of current flat-top, which corresponds to ≈ 250 additional seconds of flat-top operation that are made available.

The fusion energy output can be improved by ≈ 110 GJ ( by ≈ 55% of $W_{fus}$ for Case #001) that way. s/q is only relevant for the optimisation of $W_{fus}$ as far as total heating power should not drop down below the level where the transition to good quality H-mode could be hindered or delayed. This can happen for instance if one allows the L-H transition to take place at an earlier time during the current ramp-up phase.

Another concern for current density profiles with very small peakedness that could occur if the flux saving techniques are applied too aggressively at current ramp-up, is the internal inductance li(3), which may reach very low values that could become a problem for plasma control. In Cases ##008-010, li(3) temporarily reaches minimum values of ≈ 0.63 in the early phase of flat-top. In dedicated calculations made with CREATE-NL, it was shown, that these very low li(3) values are still manageable. The operation at a flat-top current of $I_{pl} > 17$ MA may not be possible from a PF coil current control point of view though, as the lower limit in li(3) may be violated. In the current ramp-down phase, a strong rise in li(3) to values li(3)>1.5 happens after the back-transition to L-mode in all 15-17 MA scenarios (Cases ##001, 008-010) at similar level of $I_{pl}$, which may become an issue for vertical stability control.

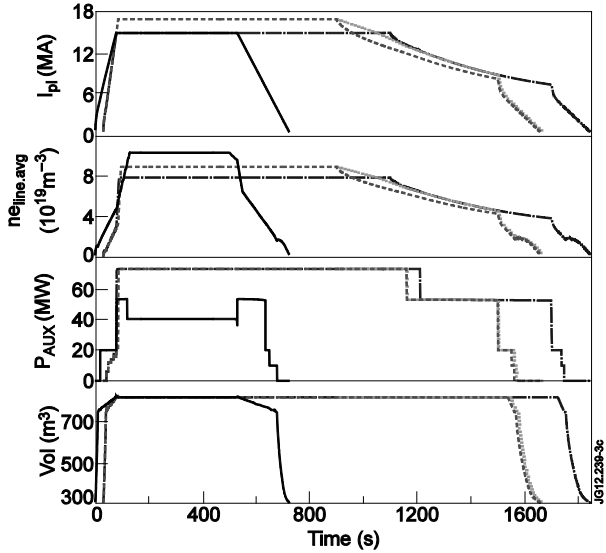

*Figure 12 – $I_{pl}$, $n_{e\ lin.\ avg.}$, $P_{AUX}$ and plasma volume for Case #001 (solid), Case #008 (chain), Case #009 (dash), and Case #010 (dotted).*

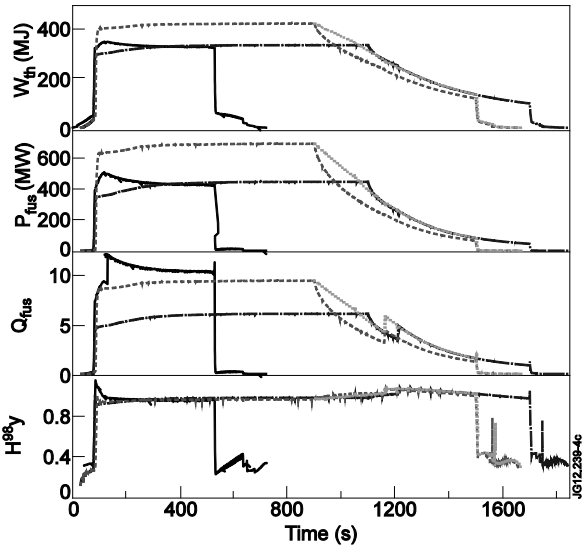

*Figure 13 – Thermal energy, $P_{fus}$, $Q_{fus}$ and $H_{98,y}$ for Case #001 (solid), Case #008 (chain), Case #009 (dash), and S3 Case #010 (dotted).*

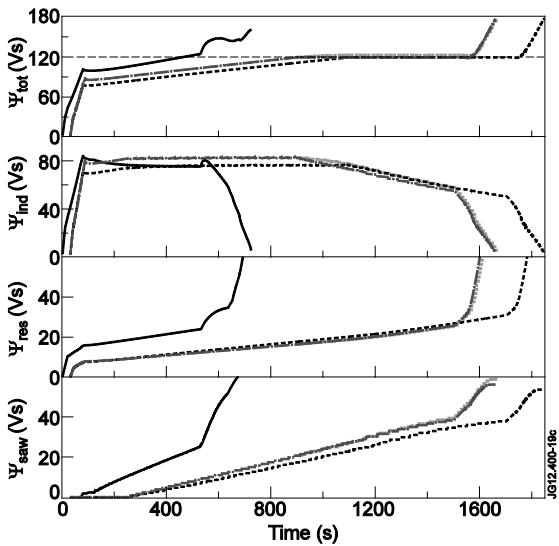

*Figure 14 $\Psi_{total}$, $\Psi_{inductive}$, $\Psi_{res}$, $\Psi_{saw}$ for Case #0010 (solid), Case #008 (chain), Case #009 (dash), and Case #010 (dotted).*

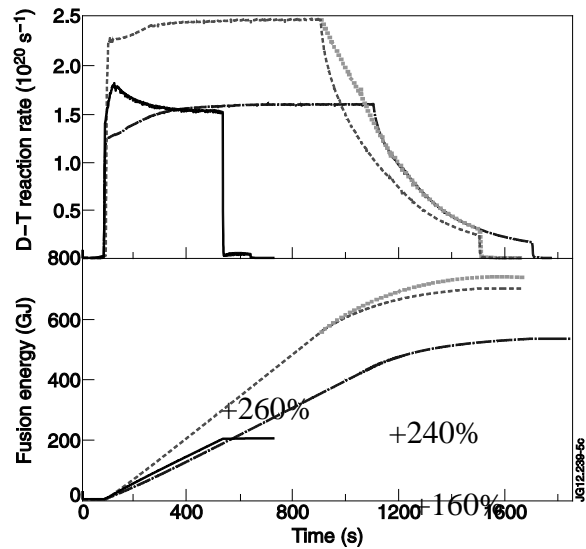

*Figure 15 Fusion reaction rate (top) and $W_{fus}$ (bottom) for Case #0010 (solid), Case #008 (chain), Case #009 (dash), and Case #010 (dotted).*

Techniques that help to minimise the rise in $\Psi_{res}$ and $\Psi_{saw}$ during the flat-top are particularly important for the optimisation wrt. $W_{fus}$. In Cases ##008-010, the decrease in density and the increase in auxiliary heating and current drive (with slightly more core-localised heat deposition) have helped considerably to keep the flux loss rates at a low level in the burning phase (≈ -45% compared to Case #001). In this period, only a moderate reduction in $n_e$ can be envisaged though that is within a range where fusion performance remains unaffected and adverse side effects such as an increase in heat flux through the separatrix caused by reduced core radiation as well as higher energies carried by the effluent particles can still be handled.
The current ramp-down phase can also contribute significantly to the maximisation of $W_{fus}$, provided that H-mode is maintained for as long as possible and flux limits are fully exploited. If $\Psi_{tot}$ is always kept at the allowed maximum for plasma shape and stability control by PF coil currents, one may expect an enhancement in $W_{fus}$ by >≈ +50%.

*In conclusion*, combining techniques for the increase in fusion energy production per discharge, the ITER baseline scenario could be optimised in a way that might permit an enhancement of $W_{fus}$ by ≈ 150-250%. It may be advantageous to operate at a higher flat-top current of $I_{pl} = 17$ MA, to increase $dI_{pl}/dt$ at ramp-up, maximise heating and current drive, operate at low to medium plasma density, to maintain H-mode for as long as possible and to stay at the maximum allowed flux level in the ramp-down phase. In the optimised scenarios that have been simulated, the flux limit has not yet been fully exploited, meaning that a further enhancement in $W_{fus}$ could be achieved by a further optimisation of the ramp-down phase.

**5. Specification of the reference Hybrid and Steady State scenarios**

The Hybrid scenario has been specified as a scenario at intermediate plasma current (with respect to the "regular H-mode" and "steady-state" scenarios at respectively Ip =15 MA and Ip = 9 MA). The motivation is to see whether a long duration discharge can be obtained with the ITER baseline heating mix at an intermediate current value, while the confinement is assumed to be improved with respect to the usual H-mode energy confinement scaling expression IPB98(y,2), as is achieved in many "hybrid scenario" experiments in present tokamaks (see, for example, [32] and references therein). From at least the MHD activity point of view, minimising the effect of sawtooth activity, for example, by maintaining $q_{min}>1$ is a key to sustain such scenarios and a key element of the scenario optimisation is to delay as much as possible the occurrence of the q = 1 surface.
A key originality of the requested scenario is to use a peaked density profile, in line with state-of-the-art physical understanding of electron transport, whereas a lot of previous ITER scenarios studies have assumed flat density profile inside the H-mode pedestal. As will be shown in the study, this has a beneficial impact on the bootstrap current and makes easier the sustainment of q above unity. Therefore the following Hybrid scenario specification/target was used:

- Plasma current $I_p$ = 12 MA
- Use ITER baseline heating mix with maximum power during the burn phase to maximize $\beta_N$
- Use fixed H98(y,2) = 1.3 for enhanced confinement during flat-top (transport coefficients scaled to follow H98(y,2) = 1.3 with a radial shape similar to the one that would be obtained with the Bohm/gyroBohm model)
- Delay as much as possible the occurrence of the q = 1 surface
- Target burn phase duration of the order of 1000 s
- Q >= 5

Regarding the Steady State (SS) scenario there are two possible venues to explore them. The first one relies on the assumption of globally improved confinement, without describing a local ITB. Its design is quite similar to the hybrid scenario presented earlier, at reduced plasma current Ip = 10 MA. At this current, assuming a globally improved confinement factor of H98 = 1.35 and making use of the maximum power available in the baseline ITER heating mix plus adding 15 MW of LHCD allows reaching fully non-inductive current drive and steady-state conditions. Conversely to the second version of the steady-state scenario with ITB, the heating and current drive in this scenario does not require detailed tailoring and its sole function is to provide fully non-inductive current drive. 3000 s of plasma have been simulated and shows that steady-state conditions are indeed reached within this duration. The resulting q-profile is slightly hollow with $q_{min}$ below 2. The scenario uses a peaked density profile.

The second choice relies on the assumption of improved confinement via the creation and sustainment of an Internal Transport Barrier due to negative magnetic shear. The approach and the transport model used are similar to the ones described in [33], although the scenario has been redeveloped to take into account a peaked density profile.

The following methodology was used for both hybrid and SS scenarios. First, the scenario is simulated using CRONOS only, optimizing the plasma current waveform and the heating & current drive schemes in order to meet the scenario specifications. CRONOS is used to solve self-consistently current diffusion, electron and ion heat transport, calculations of all source terms of these equations, and the equilibrium with a prescribed time-dependent plasma boundary. Density and $Z_{eff}$ profiles are prescribed; toroidal rotation is not taken into account in the calculations. The plasma boundary evolution is prescribed from the CREATE simulations of the current ramp-up done for Case#001 (H-mode 15 MA), assuming the shape being given as a function of the plasma current. The result of this first work is to obtain optimized conditions for reference Hybrid and Steady-State scenarios.

The compatibility of the obtained scenario with the PF systems is then assessed with two DINA-CH/CRONOS coupling methods. First, the "Prescribed CRONOS" mode is run, in which DINA-CH is run solving for current diffusion and free boundary equilibrium evolution using the CRONOS results in terms of kinetic profiles as prescribed. This phase enables us to simulate the free-boundary evolution of the plasma in a fast and efficient manner, enabling the development of an appropriate plasma shape waveform, plasma current waveform, and PF coil currents waveforms with respect to ITER capabilities. This phase also allowed us to develop and test our plasma control strategy for the hybrid and steady-state pulses. Finally, the "Self-consistent" mode is run, where DINA-CH and CRONOS are run in coupled mode, resulting in a self consistent evolution of the free boundary equilibrium, current and kinetic profiles, which are evaluated and exchanged between codes on every time step (5ms).

The MHD stability of the established scenarios is then checked a posteriori with the MISHKA code.

## 6. The main results of predictive modelling of Hybrid and Steady State scenarios

*Hybrid scenario results.*

A scenario with all the requested characteristics has been established. The general specification and results are shown in table 3. Regarding heat transport, Bohm/gyroBohm model is used during the L-mode phases, in its original L-mode version without magnetic shear dependence [18] whereas a fixed H98(y,2) = 1.3 is used as a enhanced confinement during the flat-top phase. Transport coefficients are scaled to follow H98(y,2) = 1.3, with a radial shape similar to the one that would be obtained with the Bohm/gyroBohm model. Toroidal rotation in ITER plasmas is expected to be low due to the low external torque generated by the NBI heating system [15]. Therefore in these simulations toroidal rotation has not been taken into account.

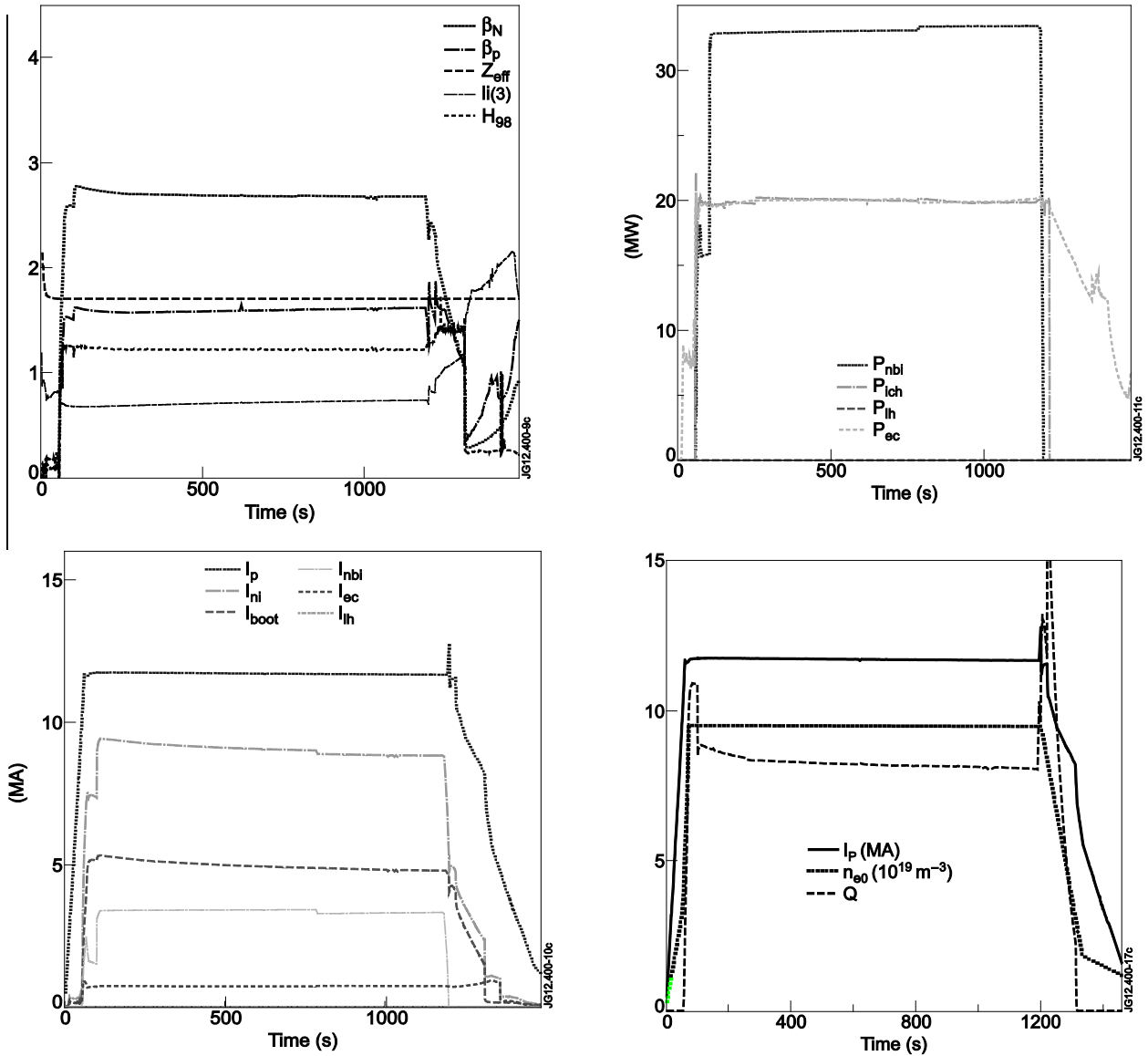

*Figure 16. Time evolution for the ITER hybrid scenario (CRONOS only). Top left: $\beta_N$, $\beta_p$, q95, li(3) and H98. Top right: NBI, ICRH, LH, and ECRH powers. Bottom left: Total ($I_p$), non-inductive ($I_{ni}$), bootstrap ($I_{boot}$), NBI ($I_{NBI}$), ECCD ($I_{EC}$) and LH currents. Bottom right: Total current (Ip), central density ne(0) and fusion gain Q.*

As expected, the scenario is nor steady-state (about 25 % of the current is driven inductively) neither stationary. Owing to the high electron temperature (central Te(0) above 30 keV), the current profile diffuses over the whole duration of the burn phase. With the ITER baseline heating mix and assuming purely neoclassical resistive current diffusion (no q-profile modification by possible MHD effects), the safety factor slowly drops towards unity and would eventually reach it if the burn phase would be extended significantly beyond 1000 s. The key to scenario optimization has been to delay as much as possible the decrease of the safety factor, by acting on a number of parameters.

− A separatrix leading to q95=4.3 at Ip=12MA has been chosen. The evolution of q has been shown to be very sensitive to q95 as for q95<4 the sustainment of q for times longer than 1000s is an issue.

− The current ramp-up rate is dIp/dt = 0.18 MA/s which leads to Ip=12MA at t=60s. No faster or slower ramp-up rate compared to the typical H-mode scenario has been considered. 8 MW of ECRH/ECCD from the ITER top launchers are used after the X-point transition. The heating applied is necessary to mildly reverse the q profile and slow-down its evolution during the ramp-up.

- The main heating phase is started at Ip = 10 MA during ramp-up with the aim of slowing-down the q profile evolution when its minimum is about 1.5.

- The L-H and H-L transitions are described by an immediate change of the transport coefficient model as the LH power threshold is reached. The appearance / disappearance of the density pedestal is prescribed at the same time. No "Type III H-mode" –like transition is used, i.e. the transport model switches directly from L-mode to high performance Type I H-mode (and vice-versa). The temperature dynamics is calculated self-consistently following the heat transport equations. $\rho_{ped} \approx 0.95$, $n_{ped} \approx 0.55\ 10^{20}$ m-3, $T_{ped} \approx 4.5$ keV. This corresponds to a pedestal pressure of 75kPa, which follows the EPED model for the scenario considered. We have verified a posteriori that this pedestal is linearly MHD stable.

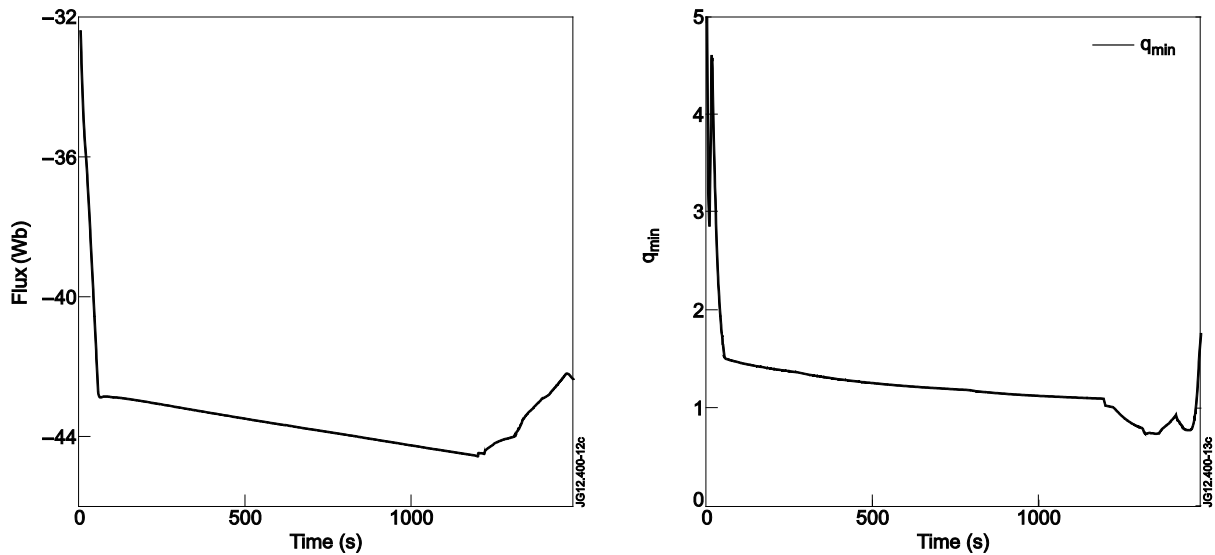

*Figure 17: Time evolution for the ITER hybrid scenario (CRONOS only). left: Flux consumption. Right: Minimum of the safety factor profile as a function of time.*

− Most ITER scenario studies up to now have considered flat density profiles inside the pedestal. In the past decade, both experimental evidence and theoretical considerations have shown that a significant electron density peaking is most likely to occur in ITER plasmas, owing to the low collisionality since density peaking tends to increase with decreasing Greenwald fraction and collisionality [34]. Since collisionality is very low and the Greenwald fraction considered for this scenario is relatively high, a mild peaking factor of $n_e(0)/n_e(0.8)=1.4$ is prescribed which has a beneficial effect (with respect to flat density profile) on the amount of bootstrap current which highly contributes to slow-down the evolution of q.
− The ITER baseline heating mix with maximum power during the burn phase: an optimized use has been elaborated in order to delay the occurrence of the q = 1 surface. The NBI power has been split into 16.5MW on-axis and 16.5MW off-axis, in order to avoid a current hole due to the excess of off-axis current. 20 MW of ECRH/ECCD from the equatorial launchers are used. With this configuration the deposition is obtained at r=0.4 which has a strong impact on shaping the q profile. Finally, 20MW of ICRH are also used with standard configuration.

With all these elements, a burn phase duration of the order of 1000 s has been achieved in the simulations and could be a priori extended since the flux consumption limit has not been reached yet. The (self-imposed) limit to the burn duration is given by the slow evolution of the q profile that would reach 1 after 1000 s. The main characteristics of the optimised Hybrid scenario are shown on figures 16 and the flux consumption and evolution of the minimum q in figure 17. The global fusion gain is Q =7.8 with $\beta_N$=2.65 and $\beta_p$=1.45. The currents obtained are the following: $I_{boot}$=4.7MA ($f_{boot}$=39%), $I_{NBCD}$=3.5MA ($f_{NBCD}$=29%), $I_{ECCD}$=0.74MA ($f_{ECCD}$=6.1%), for a total non-inductive fraction of ~75%.

*Sensitivity analysis for Hybrid scenario*

As previously shown, the hybrid scenario has a q profile in the core with a minimum value above 1 for the full simulation (although not in stationary conditions): the feature, which is highly desirable for the hybrid regimes and obtained in many present day experiments. However this feature highly depends on the assumptions assumed in this paper. Several sensitivity analyses have been carried out: the separatrix shape, the density peaking, the value of H98(y,2) and the heating during the ramp-up. As shown in [35], small changes on these parameters can modify the time of occurrence of q=1 surface and the q profile shape. A simultaneous modification of these parameters would lead to a serious deterioration of the scenario, reducing the duration until which $q_{min}$ eventually drops below 1.
During the coupled DINA-CH/CRONOS simulation it also appeared that one of the most outstanding issue consisted of constraining the q-profile evolution. Our approach was a 'fire and forget' method, in the sense that there was no feedback on the q-profile and that the heating mix prepared before the pulse was not updated during the discharge. This method was successful, but it required considerable effort and several iterations, since the DINA-CH&CRONOS self-consistent simulation of the hybrid scenario unveiled the recurring appearance of a current hole at the centre of the plasma. This phenomenon required multiple modifications to the nominal heating scheme developed using CRONOS. It is also worth noting that since the Hybrid scenario uses a lower level of plasma current (and energy content), a fast transient phenomenon, such as the L-H or H-L transition, did not push the PF control system beyond its limit. Our control scheme was sufficient to prevent the plasma from being limited outboard during the L-H transient as expected. However, the H-L transition is more difficult and the plasma was slightly further away from the wall when it occurred, thereby avoiding wall contact for this specific simulation. Further tests are underway to explore ways of improving this limiting condition.

*Results for the Steady State scenario without Internal Transport Barrier (ITB).*

This scenario relies on the assumption of globally improved confinement, without describing a local ITB. Its design is quite similar to the hybrid scenario presented above, at reduced plasma current Ip = 10 MA. At this current, assuming a globally improved confinement factor of H98 = 1.35 and making use of the maximum power available in the baseline ITER heating mix plus adding 15 MW of LHCD with wave frequency 5 GHz allows reaching fully non-inductive current drive and steady-state conditions. The main characteristics for the scenario without internal transport barrier can be found in table 3. The steady-state scenario is simulated during 3300s. The currents obtained are the following: $I_{boot}$=5 MA ($f_{boot}$=50%), $I_{NBCD}$=3.6MA ($f_{NBCD}$=36%), $I_{ECCD}$=0.7MA ($f_{ECCD}$=7%), $I_{LH}$=0.7MA ($f_{LH}$=7%) with Q ~ 5.0 and $\beta_N$= 2.70, $\beta_p$= 1.66.

More details on the results for the hybrid and steady-state with no ITB scenarios, as well as a new series of self-consistent CRONOS/DINA-CH simulations will be presented in a recently submitted paper dedicated to these advanced scenarios [35].

*Results for the Steady State scenario with ITB.*

This scenario is based on the steady-state scenario analyzed in [36] in which the flat density profile has been changed by a peaked one with peaking factor $n_e(0)/\langle n_e \rangle$ = 1.4. Unlike the previous scenarios without ITB, here the confinement factor is not fixed but obtained from the transport model [37]: $\chi_i = \chi_e = \chi_{i,neo}+0.4(1+3\rho_2)F(s)$, where $\rho$ is the normalized radius coordinate, and $F(s)=1/(1+exp(1-s))$ with s the magnetic shear. This model has been extensively used to establish reference ITER scenarios with ITB and it is mainly based on the experimental results obtained in JT-60U [38]. A word of caution is needed before we start discussing the results of our analysis for this scenario. There are two main uncertainties in projecting this scenario to ITER and both relate to the fact that the mechanisms of the ITB formation and sustainment are not well understood. First of all, getting steady state plasma in ITER requires strong contribution from the bootstrap current, which relies a on very good confinement with H98(y,2)≥1.7. Improved confinement with H98(y,2)=1.7 has been already obtained under ITER relevant conditions in the JT-60U tokamak [39] by developing strong ITB's with reversed magnetic shear. The steady-state scenario found in this paper highly resembles those regimes. Scenarios with ITB have been also obtained in the JET tokamak with H98(y,2)=1.5 [40], although their sustainability was more limited and their performance was lower compared to JT-60U ITB plasmas. The second problem is that both ITB plasmas in JT-60U and JET were not truly steady state and lose their ITB and good confinement after some time mainly due to slow evolution of the q-profile. Therefore, the reliable extrapolation of such regimes to ITER requires a more detailed analysis than the one performed here. We therefore consider our results in this paragraph as indicative rather than conclusive.

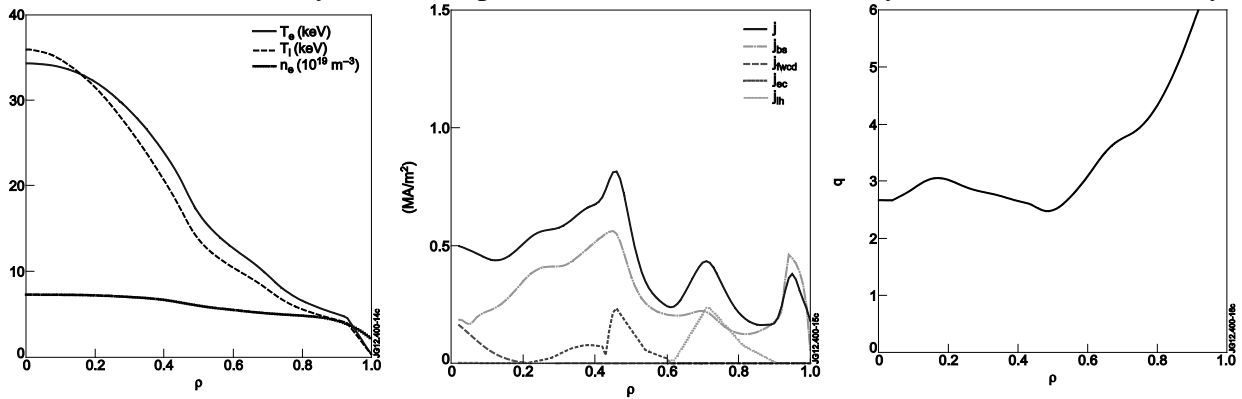

*Figure 18: Profiles during the burn phase of ITER steady-state scenario with ITB. Left: Electron temperature Te, Ion Temperature Ti and electron density ne profile. Middle: Current density of bootstrap (chain), fast wave (long dash), Lower Hybrid waves (dotted), Electron Cyclotron current drive (dashed), total (solid). Right: Safety factor profile.*

The transport model assumption of an ITB based on negative magnetic shear makes the ITB sustainment delicate and requires careful tailoring of the current profile during the burn phase. This is the well known "current alignment" issue. While the high confinement with H98y=1.7 is needed to ensure enough bootstrap current, generated by the strong pressure gradient just inside the ITB, some extra measures are needed to ensure the stability of the current profile inside ITB. Indeed it is known, that strong bootstrap current within ITB increases the local magnetic shear. In turn, this weakens the ITB locally while moving it inwards. This process is at the origin of the shrinking of the ITB and a reduction of the fusion performance.

In order to prevent the shrinking of the ITB, one needs to locally freeze the magnetic shear at a low value, so that the small q-profile modification induced by the ITB is not sufficient anymore to change the ITB properties. Freezing the magnetic shear can be obtained in ITER by using 20 MW of ECCD around mid-radius. The ITB on electrons and ions can be therefore be obtained at mid radius as shown in Figure 18.

In addition to this local control of the magnetic shear, the whole plasma current has to be sustained non-inductively. Using NBCD is detrimental for this scenario because it would drive current inside the ITB and thus make it shrink. Therefore 20 MW LHCD have been added with respect to the ITER baseline heating mix to drive the necessary non-inductive current off-axis (beyond mid-radius). The LH current drive being located outside of the ITB, it does not perturb it as shown in Figure 18. Finally, 20 MW of ICRH are also used to have sufficient pressure and bootstrap current. This carefully tailored current profile is well aligned with the ITB, which is shown to be fully steady-state in the presented 1000 s burn phase simulation.

The magnetic shear is not strongly negative, something that can prevent deleterious MHD phenomena.

The main characteristics for the scenario with internal transport barrier can be found in table 3. This steady-state scenario is simulated for 1000 s. The currents obtained are the following: $I_{boot}$=6.1MA ($f_{boot}$=76%), $I_{ECCD}$=0.6MA ($f_{ECCD}$=7.5%), $I_{LH}$=0.9MA ($f_{LH}$=11.2%), $I_{Ohm}$=0.4MA ($f_{Ohm}$=5.0%) with Q ~ 5, $\beta_N$=2.75, $\beta_p$=2.35. The peaked density has a beneficial effect on the amount of bootstrap current and indirectly on the LHCD efficiency (lower edge density).

## 7. Conclusions

– ITER baseline 15 MA inductive burn scenario and its variants (including scenarios in H and He also at reduced current and toroidal field) have been self-consistently modelled using link between CREATE-NL 2D equilibrium code and JINTRAC 1.5D transport code;
– Simulation of 15MA DT baseline finds compatibility with present PCS design;
– Q=10 operation is predicted, but sensitive to ETB assumptions;
– A slow Ip ramp-up with late L-H is found to optimise fusion power, whereas a fast Ip ramp-up and early L-H optimises pulse length;
– Sensitivity study allows optimisation of the baseline inductive burn scenario
– Variable pellet frequency is required to achieve the desired density waveform
– Studies of L-H and H-L transitions suggests PCS can control plasma, but near or outside limits for some H-L cases.
– Optimised inductive burn scenario with 17 MA flat top current has been designed and simulated with the total fusion energy production per discharge enhanced wrt 15MA baseline inductive burn scenario (Case#001) by ≈ 150-250%.
– In the hybrid scenario, the baseline heating and current drive mix has been optimized to delay as much a possible the occurrence of the q = 1 profile, assuming purely neoclassical resistive current diffusion. Under the modelling assumption used (essentially, prescribing

- a finite density profile peaking and predicting current and temperatures using a transport model yielding an H-mode energy confinement enhancement factor of $H_{98} = 1.3$), at $I_p = 12$ MA ($q_{95} = 4.3$) a scenario where the safety factor q remains above unity during 1000 s of burn phase has been obtained.
- This scenario uses the ITER baseline heating mix with maximum power during the burn phase. The 20MW of ECCD are essential to delay the occurrence of $q = 1$ beyond 1000 s.
- Two steady-state scenarios with all the requested characteristics have been established The first one relies on the assumption of a globally improved confinement and it is therefore much less sensitive, though the physics basis for such an enhanced confinement is not specified.. The second one, involving an Internal Transport Barrier based on magnetic shear requires a careful tuning of the current profile and has thus a narrow operational window.
- Both steady-state scenarios involve the use of additional 15 to 20 MW LHCD with respect to the ITER baseline heating mix, which is used to drive non inductive current far off-axis. This fulfils two purposes : i) drive sufficient additional current to reach fully non-inductive current drive and ii) avoid driving strong current inside mid-radius which would be detrimental to the sustainment of an ITB based on negative magnetic shear.

*This work was partly funded by the ITER Organisation and F4E under grant G255. The views and opinions expressed herein do not necessarily reflect those of the F4E or those of the ITER Organization.*

**Table 1 - 0D parameters for the Case #001 scenario.**

| | |
|---|---|
| Plasma major radius $R_0$ | ≈ 6.2 m |
| Plasma minor radius $a_0$ | ≈ 2.0 m |
| Toroidal field strength at $R_0$ | ≈ 5.3 T |
| Plasma fuel | 1:1 D-T mixture |
| Flat-top plasma current | ≈ 15 MA |
| Flat-top Greenwald fraction | ≈ 80-85% |
| Ballooning stability parameter $\alpha_c$ | ≈ 1.7-1.9 |
| ETB width on outer mid-plane | ≈ 6-8 cm |

**Table 2 – Heating scheme, plasma current and confinement state for the Case #001 scenario.**

| t [s] | ECRH [MW] | ICRH [MW] | NBI [MW] | Ipl [MA] | Conf. state |
|---|---|---|---|---|---|
| 1.3-20 | - | - | - | 0.5-5.4 | L-mode |
| 20-80 | 20 on-axis | - | - | 5.4-15 | L-mode |
| 80-120 | - | 20 | 33 | 15 | H-mode |
| 120-530 | - | 7 | 33 | 15 | H-mode |
| 530-630 | - | 20 | 33 | 15-7.5 | L-mode |
| 630-650 | 10 | 10 | - | 7.5-6.0 | L-mode |
| 650-700 | 10 | - | - | 6.0-2.2 | L-mode |
| 700-723.3 | - | - | - | 2.2-0.5 | L-mode |

**Table 3. Main characteristics of the advanced scenarios analyzed.**

| | **ITER hybrid** | **ITER steady-state (No ITB)** | **ITER steady-state (ITB)** |
|---|---|---|---|
| **Ip (MA)** | 12 | 10 | 8 |
| **Bt (T)** | 5.3 | 5.3 | 5.3 |
| **$q_{95}$** | 4.3 | 4.8 | 6 |
| **κ/δ** | 1.8/0.4 | 1.8/0.4 | 1.9/0.5 |
| **Total $\beta_N/\beta_p$** | 2.65/1.45 | 2.70/1.66 | 2.75/2.35 |
| **$f_{Gw}$** | 0.8 | 0.8 | 0.9 |
| **Pped (kPa)** | 75 | 60 | 50 |
| **$H_{98}(y,2)$** | 1.30 | 1.35 | 1.7 |
| **$P_{NBI}$(MW)** | 33 | 33 | 0 |
| **$P_{ECRH}$(MW)** | 20 | 20 | 20 |
| **$P_{ICRH}$(MW)** | 20 | 20 | 20 |
| **$P_{LH}$(MW)** | 0 | 15 | 20 |